\documentclass[aps,showpacs,prl,twocolumn]{revtex4}
\usepackage{psfig}

\begin{document}

\title{Structurally specific thermal fluctuations identify functional
sites for DNA transcription}

\author{G. Kalosakas}  \author{K.~\O.~Rasmussen}   \author{A.~R.~Bishop}
\affiliation{Theoretical Division and Center for Nonlinear Studies,
  Los Alamos National Laboratory, Los Alamos, New Mexico 87545}

\author{C. H. Choi}  \author{A. Usheva}
\affiliation{Endocrinology, Beth Israel Deaconess Medical Center,
Department of Medicine,  Harvard Medical School, 99 Brooklin Ave.,
Boston, Massachusetts 02215} 


\begin{abstract}

We report results showing that thermally-induced openings of double
stranded DNA coincide with the location of functionally relevant
sites for transcription. Investigating both viral and bacterial DNA gene
promoter segments, we found that the most probable opening occurs at
the transcription start site. Minor openings appear to be related to
other regulatory sites. Our results suggest that coherent thermal
fluctuations play an important role in the initiation of
transcription. Essential elements of the dynamics, in addition to
sequence specificity, are nonlinearity and entropy, provided by
local base-pair constraints.

\end{abstract}
\maketitle

\vspace{-0.5cm}

One of the most challenging subjects in biophysics is the relation
between biomolecular motions and function \cite{hans}.
We present an example suggesting that functionality arises
from structurally coherent dynamics, with essential
ingredients of: {\it sequence-specificity, nonlinearity and entropy};
the nonlinearity form {\it local constraints} is crucial.
In particular, remarkably successful comparisons of numerical
simulations of a minimal model (see below) of transverse dynamics
for gene promoter DNA segments with {\it in vitro} transcriptional
experiments, shows that the combination of {\it all} the above
mentioned components controls coherent ``bubble'' fluctuational
openings of base-pairs around specific sites  of promoter DNA.
Remarkably, the prominent opening occurs at the
transcription initiation site, while minor openings coincide with
regulatory sites at which transcription factors and other assisting
proteins are bound. These results demonstrate the importance of the
sequence structure, not simply as a static and passive element,
but to {\it provide the
template for specific coherent fluctuations determining function}.
These coherent structures constitute a {\it colored}
spatio-temporal stochastic environment. This is an example of the importance
of a (dynamic) landscape of substates \cite{hans}.

We have used a microscopic model proposed by Peyrard and Bishop
\cite{PB} to describe the dynamics of the openings of double stranded
DNA. This model focuses only on the most
relevant degrees of freedom, namely the transverse stretching of the
hydrogen bonds connecting complementary bases in the opposite
strands of the double helix. Its reduced character, involving
a small number of variables, makes it suitable for simulations
over relatively long times and appropriate for gathering sufficient
statistics. Subsequent key improvements \cite{DPB} succeeded in
reproducing the abrupt (first order) character of the observed DNA
denaturation transition. The potential energy of this model
reads \cite{DPB}

\begin{eqnarray}
V &=&  \sum_n  \left [ D_n \left (e^{- a_n y_n}-1\right )^2  + \right .
\nonumber \\  &&  \left .
\frac{k}{2} \left (1+\rho \exp \left [{-\beta(y_n+y_{n-1})}\right]
\right )\left (y_n-y_{n-1} \right )^2 \,\right].
\label{hl}
\end{eqnarray}

\noindent
Here the sum is over all the base-pairs of the DNA and $y_n$ denotes the
displacement from the equilibrium position of the relative distance
between the bases within the $n^{th}$ base-pair, divided over
$\sqrt{2}$. The Morse potential (other similar potentials can also be used)
in the first term provides the effective interactions between
complementary bases; it represents both the attraction due to the
hydrogen bonds forming the base-pairs and the repulsion of the
negatively charged phosphates in the backbone of the two strands
screened by the surrounding solvent. Beyond the exact details of this
interaction, an important issue is a correct description of the
nonlinearities. The parameters $D_n$ and $a_n$
of the on-site potential distinguish between the two possible
combinations of bases, i.e. adenine-thymine (A-T) or guanine-cytosine
(G-C), at site $n$, depending on the particular sequence.
The second term in the total potential energy represents the stacking
interaction potential between adjacent base-pairs. Here the nonlinear
inter-site coupling, given by the exponential term that effectively
modifies a harmonic spring constant, is essential for representing
{\it local constrains} in nucleotide motions, which result in long-range
cooperative effects \cite{DPB}. As in elastic materials \cite{MG,kerr},
it controls lattice vibrations, yielding accurate entropic terms
\cite{GLK}: the stiffening of
the coupling in the compact state compared to that in the open state
leads to an abrupt entropy-driven transition \cite{DPB}.
Physically, the constraint describes the change
of the next-neighbor stacking interaction due to the distortion
of the hydrogen bonds connecting a base-pair, mediated through
the redistribution of the electrons on the corresponding bases.

\begin{widetext}
\begin{center}
\begin{figure}
\vspace{-2cm}
\centerline{\hbox{\psfig{figure=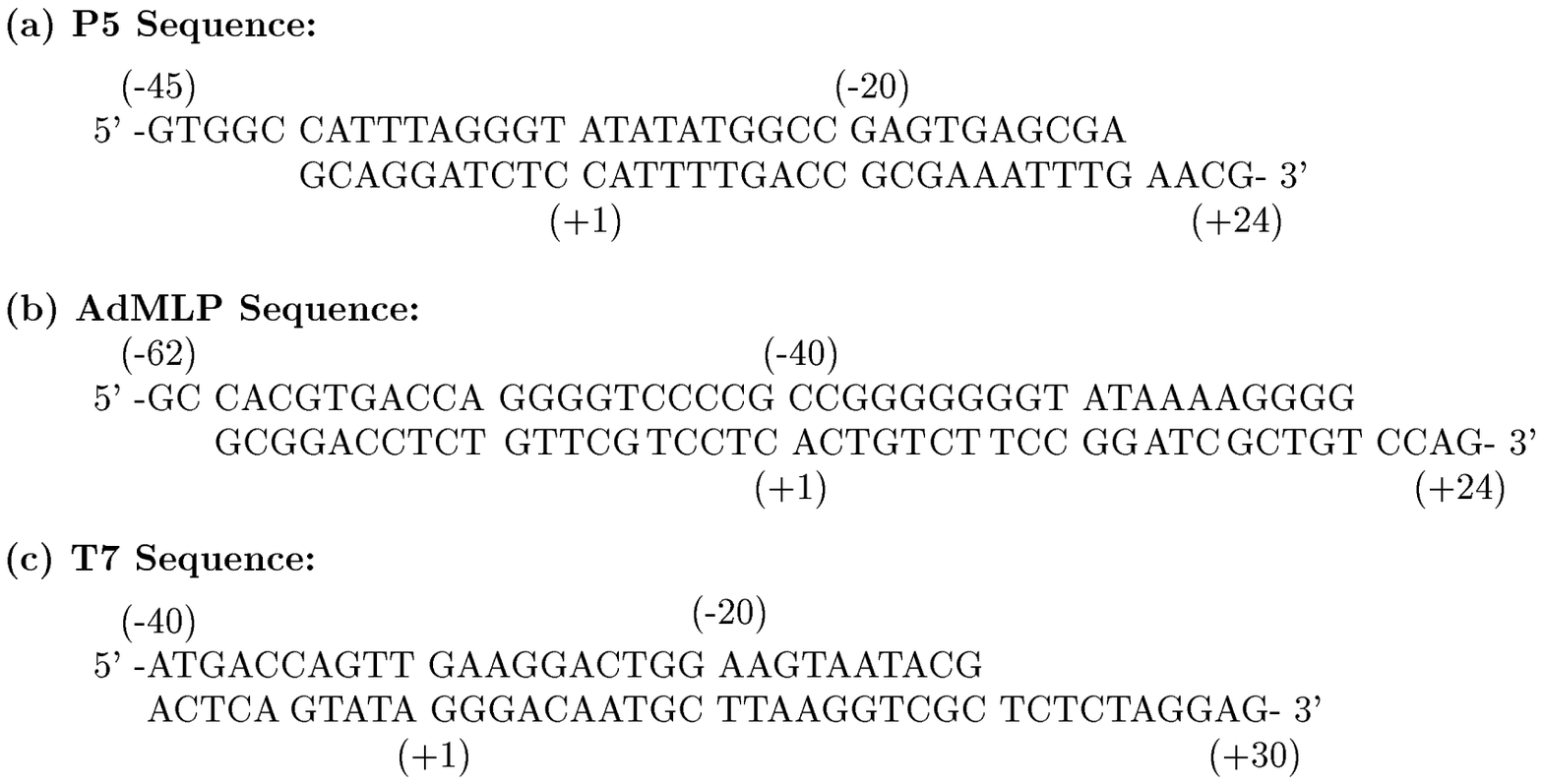,width=17.0cm,height=18.0cm,angle=0}}}
\vspace{-10.5cm}
\caption{Base-pair sequences of the studied DNA gene promoter fragments.
{\bf (a)} 69 base-pair long viral P5 promoter, {\bf (b)} 86 base-pair long
viral AdMLP promoter, and {\bf (c)} 70 base-pair long bacterial T7 promoter.}
\end{figure}
\end{center}
\end{widetext}

Despite its simplified character, this model has successfully
reproduced not only the sharp melting (denaturation) transition
occurring when the two strands of the DNA separate from each other
as temperature increases, but also the details of the
precursor (nucleation) fluctuational
openings and the dynamics upon approaching the denaturation transition.
The coexistence of two essential features are
necessary for obtaining the first-order$-$like transition
\cite{DP}: the nonlinear coupling constant that decreases in the
denaturated phase providing an increase in entropy,
and a plateau in the on-site potential which should not increase
unbounded. Regarding the precursor fluctuations, intrinsic
localized modes nucleate as nonlinear bubble opening events that
subsequently interact and grow, providing the experimentally
observed denaturation bubbles \cite{PB,englander}.
This nucleation regime, precursor to the melting,
extends over temperatures several tens of Kelvin below the
melting transition (the biologically relevant regime).

In addition to capturing the essential features of thermally induced
denaturation of long DNA chains, the model has been used
to reproduce the melting curves of very short heterogeneous
DNA segments, in excellent quantitative agreement with
experimental data \cite{CG}. Furthermore, it provides
the characteristic multi-step melting observed in single heterogeneous
DNA molecules \cite{CH}. Recently, the model has been used to
investigate charge transport properties in a flexible DNA chain,
where the charge is
coupled to the lattice degrees of freedom \cite{KKB}. The
bubbles, as relatively long-lived intrinsic inhomogeneities
(``hot-spots'') \cite{PF}, represent a colored noise environment,
which qualitatively influences charge dynamics \cite{KRB}.

Motivated by the successful description of the nonlinear thermal
fluctuations, we have applied this model to explore the possible role of the
intrinsic bubble openings for the transcriptional initiation and regulatory
sites of specific promoter DNA sequences. In particular, we have
studied the adenoassociated viral P5 promoter (P5), the adenovirus
major late promoter (AdMLP) and the bacteriophage T7 core promoter (T7).
The base-pair sequences of these promoters are presented in Fig.~1.
{\it In vitro} transcription experiments demonstrating the specific
initiation of RNA polymerase II transcription from DNA templates
containing the corresponding promoter fragments are shown in Fig.~2.
For the experimental details see reference \cite{choi}.

\begin{figure}[t]
\vspace{7cm}
\caption{Auto-radiography of [$^{32}P$]-labeled reverse transcripts
after separation by gel electrophoresis (lanes 3). Arrows indicate
the direction of specific transcription started from the initiation
site +1 in all cases. Lanes 1 indicate base position markers obtained by
chemical sequencing. {\bf (a)} P5 promoter, {\bf (b)} AdMLP promoter, and
{\bf (c)} T7 promoter. Lane 4 in (a) shows elimination of the
transcription for the mutated P5 promoter, where the nucleotides
at +2 and +3 have been changed from AT to GC. }
\end{figure}

We performed Langevin molecular
dynamics (thereby capturing thermal fluctuation and dissipation effects)
for nucleotides of mass $m$ evolving in the potential $V$ of equation
(\ref{hl}). We have used the parameter values given in reference
\cite{CG}: $k=0.025 \; eV/A^2$, $\rho=2$, $\beta=0.35 \; A^{-1}$
for the inter-site coupling, while for the Morse potential
$D_{GC}=0.075 \; eV$, $a_{GC}=6.9 \; A^{-1}$ for a GC base-pair,
and $D_{AT}=0.05 \; eV$, $a_{AT}=4.2 \; A^{-1}$ for an AT pair.
The simulated temperature was $300 \; K$ (below the melting
temperature but in the precursor regime of bubble formation).

The statistics of the thermally induced
openings was obtained using 100 different
realizations for each DNA sequence studied. We ran each realization
for $1 \; ns$, after reaching thermal equilibrium, and monitored the
state of the system every $1 \; fs$. Thus we have $10^6$ events
for each one of the 100 realizations. At every event we checked the
displacements $y_n$ of the base-pairs at each site $n$ and the following
$m-1$ ($m$ varying from 1 to 19) base-pairs. If the openings at {\it all}
these $m$ subsequent sites are greater than a threshold value
$y_{th}$ (varying from one tenth to few $A$) we assign a contribution
to the opening event at the $n^{th}$ base-pair of the sequence. The obtained
opening probabilities along the studied DNA segments
for bubble sizes of $m=10$ base-pairs and thresholds $y_{th}=1.0$
and $1.5 \; A$ for accepting an opening are presented in Fig.~3.
(We recall that the real openings are equal to $y \sqrt{2}$).
In all these cases the most probable openings are located
at the transcription start site +1 (see Fig.~2). Furthermore,
in the viral cases the other distinct openings seem to related to known
regulatory sites: in P5 the opening at the A/T rich region between
-40 and -35 corresponds to the binding site of the transcription
factor Yin Yang 1 \cite{US}, while in AdMLP the second higher opening
is close to the binding site of the TATA-box binding protein \cite{CLMW},
that is necessary for transcription. As can be seen from Fig.~3,
openings of such large widths and amplitudes are rare events in our
microscopic simulations, therefore requiring sufficient statistics.

\begin{figure}
\centerline{\hbox{\psfig{figure=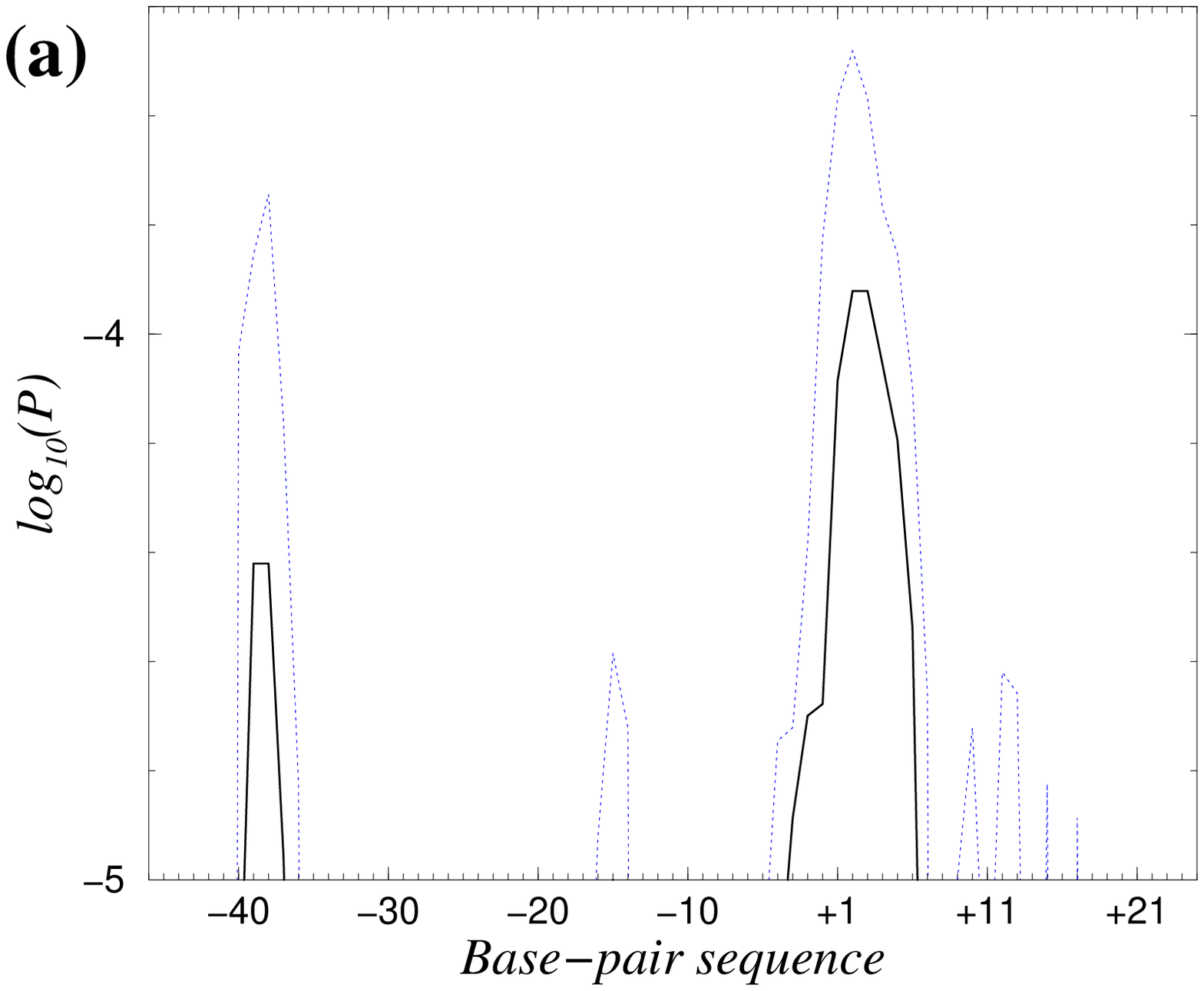,width=8.0cm,height=5.5cm,angle=0}}}
\centerline{\hbox{\psfig{figure=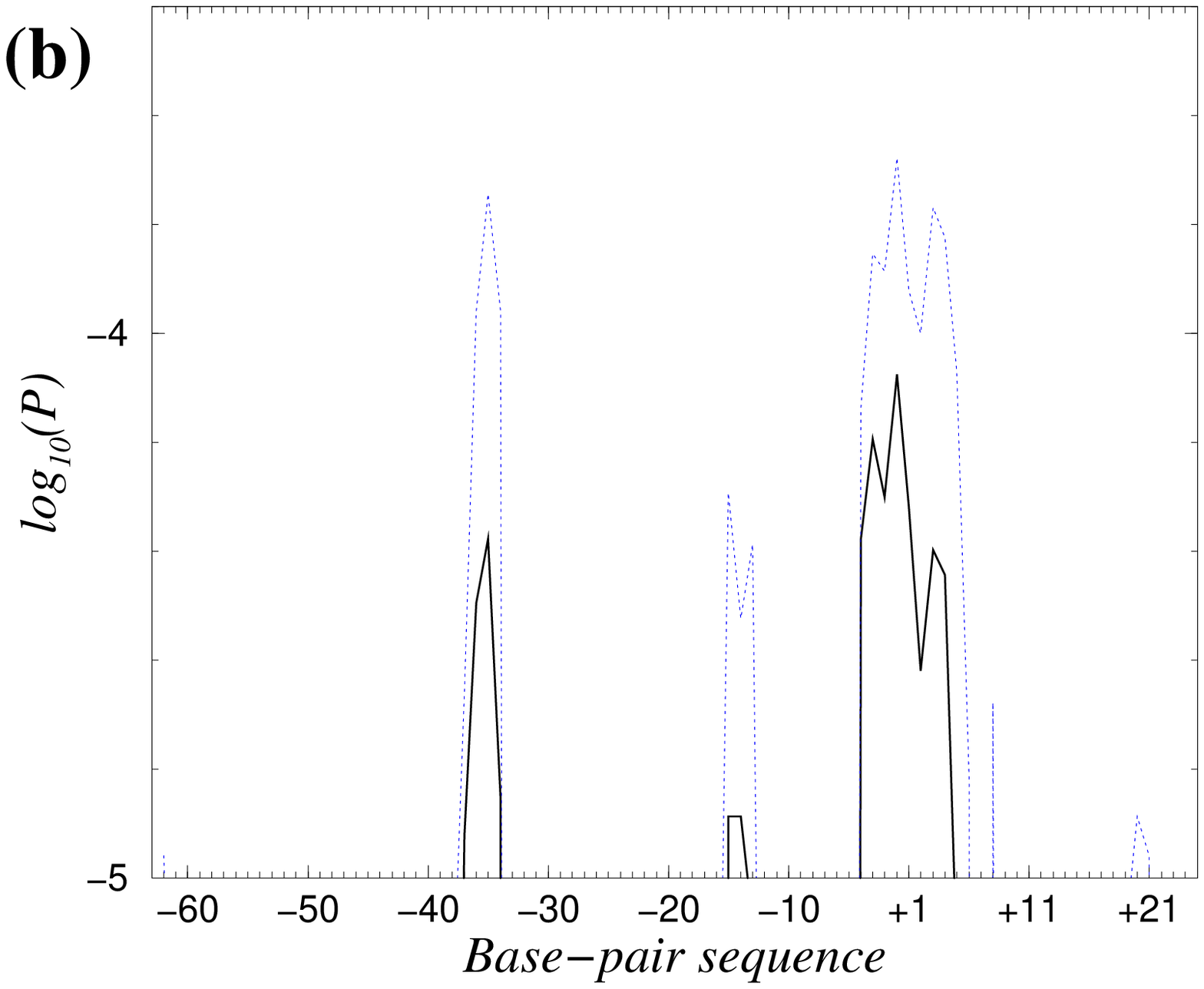,width=8.0cm,height=5.5cm,angle=0}}}
\centerline{\hbox{\psfig{figure=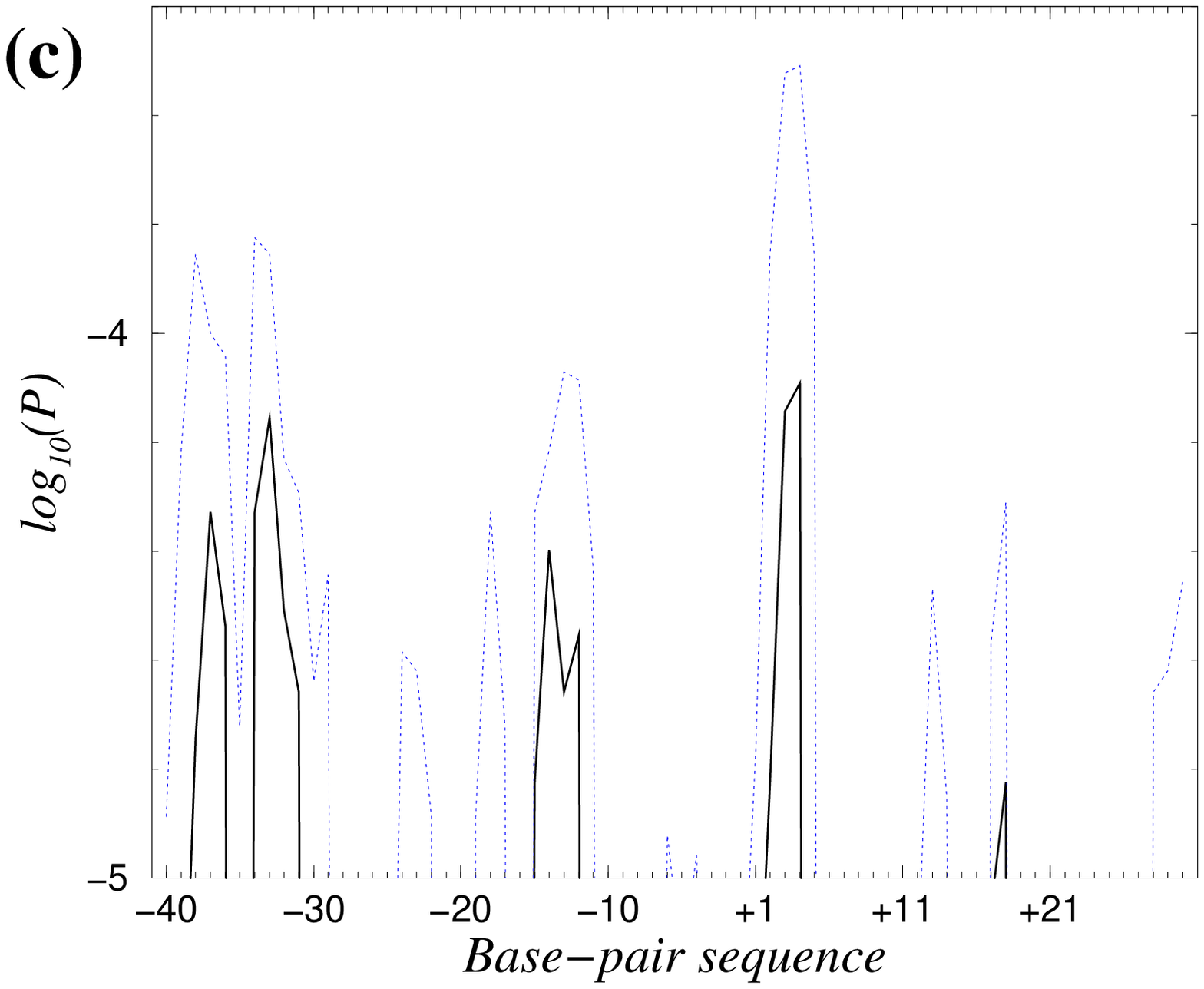,width=8.0cm,height=5.5cm,angle=0}}}
\caption{Logarithm of the probability $P$ for the occurrence of an opening
of 10 base-pair width and amplitude of more than $2.1 \; A$
(thick solid line) or $1.4 \; A$ (dotted line) starting at
a particular site $n$ of the DNA fragment, as a function of $n$, for
the {\bf (a)} P5 promoter, {\bf (b)} AdMLP promoter, and
{\bf (c)} T7 promoter.}
\end{figure}

We stress that similar local sequences do {\it not} exhibit the same
opening probabilities; equal size segments of relatively weakly
bound A/T pairs in different parts of the promoter show very
different statistics (compare for example the region around -30
with that around +1 in P5). Furthermore, the larger openings do not
occur in regions with longer A/T stretches, as might be intuitively
anticipated because of the weaker bonding. Effective long range cooperativity
(from the nonlinear inter-site potential in equation (\ref{hl})) and 
competing localization lengths due to the spatial disorder and the
nonlinearity are responsible for this high specificity: in general,
length scale competition in nonlinear systems is known \cite{SB} to
lead to complex spatio-temporal (dynamic landscape) behavior.
The sensitivity of the cooperative/competing phenomena, affected even by
a {\it single} base-pair modification, enhances the predictive power
of our model; a small mutation of the sequence (at a specific location)
is sufficient to completely eliminate bubble formation at the transcription
initiation site. In Fig.~4 we show numerical calculations of the
opening probabilities for a mutated P5 promoter, where nucleotides
+2 and +3 have been changed from AT to GC. This mutation completely
eliminates the opening at the previous transcription start site,
in agreement with no transcriptional
events occurring in the corresponding experiment (see Fig.~2a, lane 4).

\begin{figure}
\centerline{\hbox{\psfig{figure=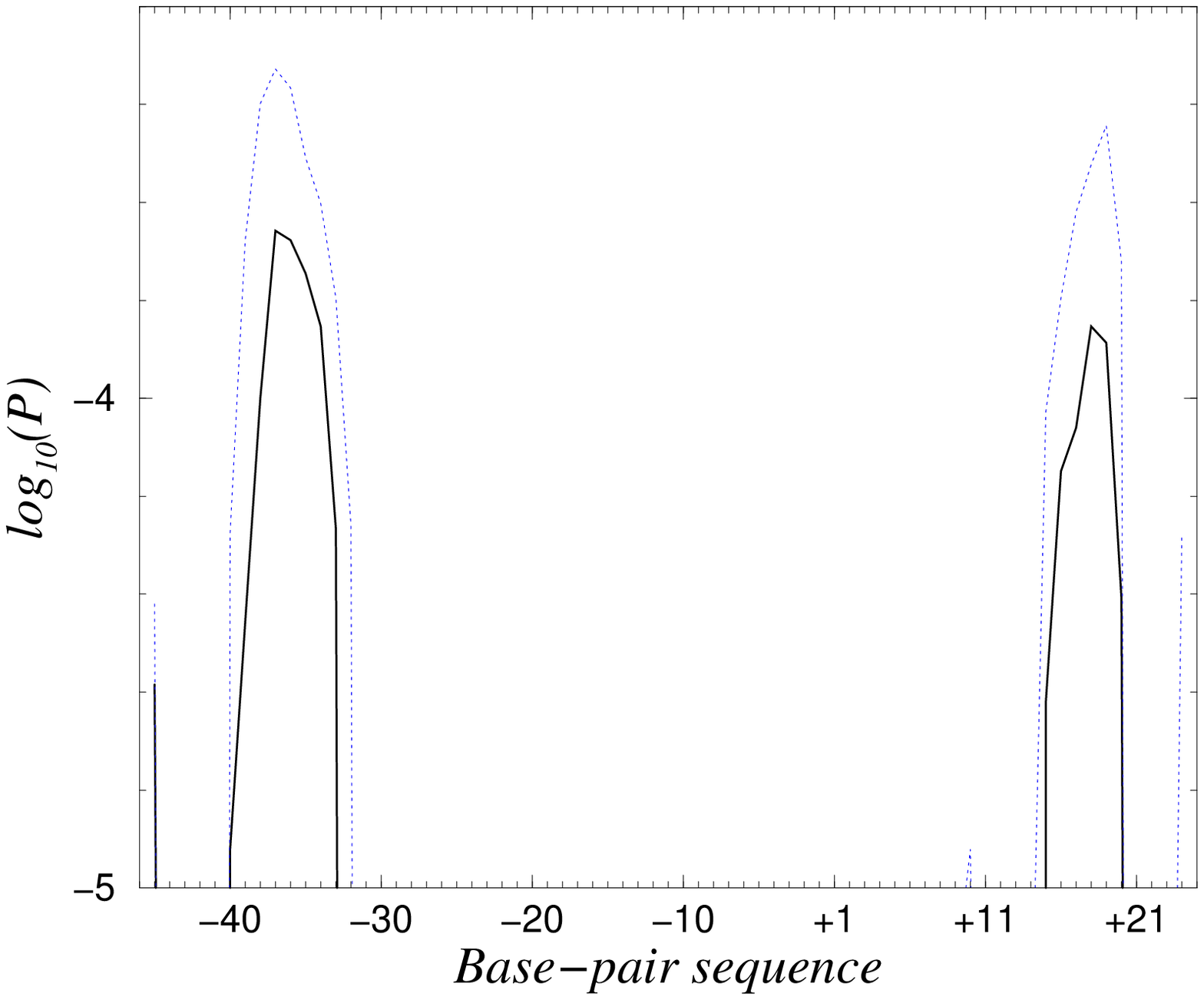,width=8.0cm,height=5.5cm,angle=0}}}
\caption{Logarithm of the probability $P$ for the occurrence of an opening of
10 base-pair width and amplitude of more than $2.1 \; A$
(thick solid line) or $1.4 \; A$ (dotted line) starting at
a particular site $n$ of the DNA fragment, as a function of $n$ for
the mutated P5 promoter (see text).}
\end{figure}

\begin{figure}
\centerline{\hbox{\psfig{figure=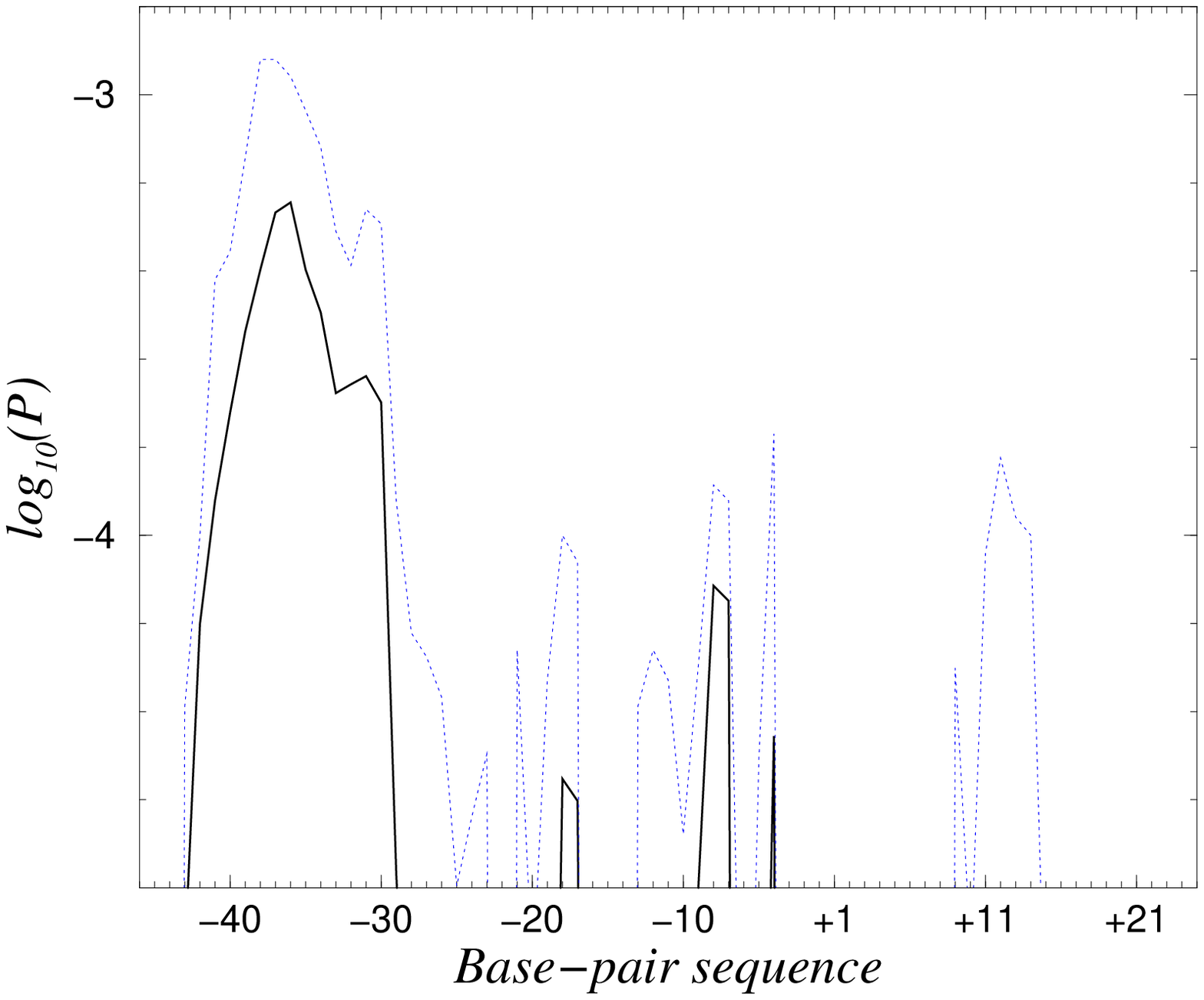,width=8.0cm,height=5.5cm,angle=0}}}
\caption{Logarithm of the probability $P$ for the occurrence of an opening of
10 base-pair width and amplitude of more than $2.1 \; A$ (thick solid
line) or $1.4 \; A$ (dotted line) starting at a particular site $n$ of
the DNA fragment, as a function of $n$ for the P5 promoter, by
linearizing the stacking interaction term of the potential
(i.e. setting $\rho=0$ in potential energy (\ref{hl})).}
\end{figure}

We emphasize that, as in previous applications of this model,
the nonlinear inter-site coupling is crucial for its success
\cite{DPB,CH}. For example, as can be seen in Fig.~5, linearizing
the stacking interaction term ($\rho=0$) results in very modified
statistics for the openings of the P5 promoter, changes the position
of the peaks along the sequence, and eliminates the successful
comparison with the experimental observations.
The nonlinear inter-site coupling constitutes
a minimal representation of the local stacking constraint between
neighboring base-pairs. As in more general situations of
displacive structural phase transitions\cite{MG,kerr},
such local constraints can lead to long-range ``elastic'' interactions
and macroscopic cooperativity.

In summary, our model and simulations suggest that structurally specific
coherent thermal fluctuations identify locations in the DNA sequences
where the RNA polymerase initiates transcription. Further, we find
indications that the thermal fluctuations also help in recruiting
other protein complexes participating in the transcriptional process,
by separating the DNA double strand at specific locations. These
bubbles precede protein binding and their possible role is limited to
the very initial steps of the transcription. This suggests that DNA,
through structurally specific dynamics, participates in directing its
own transcription.

We would like to thank H. Frauenfelder, P. Fenimore, and
J.A. Krumhansl for valuable discussions.
This research was supported by the US DoE, under
contract $W$-$7405$-$ENG$-$36$ and NIH.

\end{document}